\newcommand{\vy}[2]{#1_{\scriptscriptstyle #2}}
\documentclass[12pt,preprint]{aastex}

\shorttitle{Conditions of the Outflow in QSO~2359--1242}
\shortauthors{Korista et~al.}
\begin{document}
\title{PHYSICAL CONDITIONS IN QUASAR OUTFLOWS: \\
       VLT OBSERVATIONS OF QSO~2359--1241\altaffilmark{1}}

\author{Kirk T.\ Korista}
\affil{Department of Physics, Western Michigan University}
\affil{Kalamazoo, MI 49008-5252}
\email{kirk.korista@wmich.edu}

\author{Manuel A.\ Bautista and Nahum Arav\altaffilmark{2}} 
\affil{Department of Physics, Virginia Polytechnic and State University,
Blacksburg, VA 24061}

\author{Maxwell Moe} 
\affil{Department of Astronomy, University of Colorado, Boulder, CO}

\author{Elisa Costantini} 
\affil{SRON National Institute for Space Research} 
\affil{Sorbonnelaan 2, 3584 CA Utrecht, The Netherlands}

\author{Chris Benn} 
\affil{Isaac Newton Group, Observatorio del Rogue de los Muchachos, Spain}

\altaffiltext{1}{Based on observations made with ESO Telescopes at the Paranal Observatories under programme ID 078.B-0433(A)}
\altaffiltext{2}{previously at Department of Astronomy, University of Colorado, Boulder, CO}

\begin{abstract}
We analyze the physical conditions of the outflow seen in QSO~2359--1241
(NVSS J235953--124148), based on high resolution spectroscopic VLT
observations. This object was previously studied using Keck/HIRES
data. The main improvement over the HIRES results is our ability to
accurately determine the number density of the outflow.  For the major
absorption component, level population from five different \ion{Fe}{2}
excited level yields $\vy{n}{H}=10^{4.4}$ cm$^{-3}$ with less than 20\%
scatter. We find that the \ion{Fe}{2} absorption arises from a region with
roughly constant conditions and temperature greater than 9000~K, before
the ionization front where temperature and electron density drop. Further,
we model the observed spectra and investigate the effects of varying
gas metalicities and the spectral energy distribution of the incident
ionizing radiation field. The accurately measured column densities allow
us to determine the ionization parameter ($\log U_H \approx -2.4$) and
total column density of the outflow ($\log N_H(\rm{cm}^{-2}) \approx
20.6$). Combined with the number density finding, these are stepping
stones towards determining the mass flux and kinetic luminosity of the
outflow, and therefore its importance to AGN feedback processes.
\end{abstract}


%\objectname{J2359-1241}

\normalsize

\section{INTRODUCTION}

In recent years, the potential impact of quasar outflows on their
environment has become widely recognized  (e.g., Silk \& Rees 1998,
King 2003, Cattaneo et al.\ 2005, Hopkins et al.\ 2006).  Outflows are
detected as absorption troughs in quasar spectra that are blueshifted with
respect to the systemic redshift of their emission line counterparts.
The absorption troughs are mainly associated with UV resonance lines of
various ionic species (e.g., \ion{Mg}{2}~$\lambda\lambda$2796.35,2803.53,
\ion{Al}{3}~$\lambda\lambda$1854.72,1862.79,
\ion{C}{4}~$\lambda\lambda$1548.20,1550.77,
\ion{Si}{4}~$\lambda\lambda$1393.75,1402.77,
\ion{N}{5}~$\lambda\lambda$1238.82,1242.80).

Some quasar outflows show absorption troughs from excited and metastable
states. The ratio of the population level between the excited or
metastable states and the ground state is sensitive to the number density
and temperature of the plasma (Wampler, Chugai \& Petitjean 1995; de~Kool
et~al.\ 2001). Therefore, accurate measurements of the column densities
associated with both excited or metastable states and the ground state of
a given ion can yield the gas number density of the outflow. In addition,
these measurements and similar ones of troughs from other ions and
elements allow us to determine the ionization equilibrium and total
column density in the outflow (Arav et~al.\ 2001; Arav et~al.\ 2007).

Accurate column densities for the outflow's troughs are difficult
to determine since the outflow does not cover the emission source
homogeneously (Barlow 1997; Telfer et~al.\ 1998; Arav 1997; Arav et~al.\
2003). Over the past several years, we have developed techniques for
extracting reliable column densities for such situations (Arav et~al.\
1999a; Arav et~al.\ 1999b; de~Kool et~al.\ 2001, 2002a,b; Arav et~al.\
2002; Scott et al.\ (2004) Gabel et al.\ (2005); Arav et~al.\ 2005). These
efforts culminated with the analysis of spectroscopic VLT observations
of the outflow seen in QSO~2359--1241 (Arav et~al.\ 2008; hereafter
Paper~I). These data contain absorption troughs from five resonance
\ion{Fe}{2} lines, as well as those from several other metal species
and metastable excited state \ion{He}{1}, with a resolution of $\sim$7
km~s$^{-1}$ and signal-to-noise ratio per resolution element of order 100.

QSO~2359--1241 (NVSS J235953--124148; $E = 15.8$) is an intrinsically
reddened ($A_V \approx 0.5$), luminous ($M_B = -28.7$), radio-moderate,
optically polarized ($\sim 5\%$), low-ionization broad absorption line
quasar, at relatively low redshift $z \approx 0.868$. See Brotherton
et~al.\ (2001) for further details. Brotherton et~al.\ (2005) describes
its X-ray spectrum. An initial investigation of the physical properties
of the outflow in this object using $HST$ FOC and especially Keck HIRES
spectra is described in Arav et~al.\ (2001).

The VLT spectral data set of QSO~2359--1241 is described in detail in
Paper~I. Its unprecedented high-quality allowed us to test a variety
of absorber distribution models needed to derive reliable ionic column
densities of the outflow (see Paper~I). In the present paper we report
these column densities and use them to determine the physical conditions
within the main component of the outflow ({\bf e}, see Paper~I): the
ionization equilibrium, total column density and number density of the
absorbing material. To do so we use the photoionization code Cloudy
(Ferland et al.\ 1998) as well as a separate \ion{Fe}{2} ion model
(Bautista \& Pradhan 1998).

The plan of the paper is as follows: In Section~2 we describe the column
density measurements. In Section~3 we determine the physical conditions
in the main component of the outflow. Finally, in Section~4 we summarize
and discuss our results and provide a simple estimate of the outflow's
distance from the central continuum source.

%%%
\clearpage
\begin{deluxetable}{lrlrrrr}
\tabletypesize{\scriptsize}
\tablecaption{Absorption lines identified in the VLT spectrum
of QSO~2359--1241}  
\tablewidth{0pt}
\tablehead{
\colhead{
$\lambda$}&\colhead{$\log(gf)$\tablenotemark{a}}& \colhead{Ion}&\colhead{E$_{low}$(cm$^{-1}$)}&
\colhead{$g_{low}$}&\colhead{E$_{up}$(cm$^{-1}$)}&
\colhead{$g_{up}$}}    
\startdata 
2764.62 &-1.95  &\ion{He}{1}*  &159856 & 3& 196027 & 9\cr
2829.92 &-1.74  &\ion{He}{1}*  &159856 & 3& 195193 & 9\cr
2945.98 &-1.58  &\ion{He}{1}*  &159856 & 3& 193801 & 9\cr
3188.69 &-1.16  &\ion{He}{1}*  &159856 & 3& 191217 & 9\cr
3889.80 &-0.72  &\ion{He}{1}*  &159856 & 3& 185565 & 9\cr
2852.97 & 0.270 &\ion{Mg}{1}   &     0 & 1&  35051 & 3\cr
2796.36 & 0.100 &\ion{Mg}{2}   &     0 & 2&  35761 & 4\cr
2803.54 &-0.210 &\ion{Mg}{2}   &     0 & 2&  35669 & 2\cr
1854.72 & 0.060 &\ion{Al}{3}  &     0 & 2&  53917 & 4\cr
1862.79 &-0.240 &\ion{Al}{3}  &     0 & 2&  53683 & 2\cr
1808.01 &-2.100 &\ion{Si}{2}   &     0  &2  &55309  &4\cr
1816.93 &-1.840 &\ion{Si}{2}m* &   287  &4  &55325  &6\cr
3934.83 & 0.134 &\ion{Ca}{2}   &     0  &2  &25414  &4\cr
3969.65 &-0.166 &\ion{Ca}{2}   &     0  &2  &25192  &2\cr
2576.87 & 0.433 &\ion{Mn}{2}   &     0  &7  &38807  &9\cr
2594.49 & 0.270 &\ion{Mn}{2}   &     0  &7  &38543  &7\cr
2606.46 & 0.140 &\ion{Mn}{2}   &     0  &7  &38366  &5\cr
2344.2139&  0.057& \ion{Fe}{2}  &      0& 10&  42658&  8\cr
2374.4612& -0.504& \ion{Fe}{2}  &      0& 10&  42115& 10\cr
2382.7652&  0.505& \ion{Fe}{2}  &      0& 10&  41968& 12\cr
2586.6500& -0.161& \ion{Fe}{2}  &      0& 10&  38660&  8\cr
2600.1729&  0.378& \ion{Fe}{2}  &      0& 10&  38459& 10\cr
2333.5156& -0.206& \ion{Fe}{2}*&    385&  8&  43239&  6\cr
2365.5518& -0.402& \ion{Fe}{2}*&    385&  8&  42658&  8\cr
2389.3582& -0.180& \ion{Fe}{2}*&    385&  8&  42237&  8\cr
2396.3559&  0.362& \ion{Fe}{2}*&    385&  8&  42115& 10\cr
2599.1465& -0.063& \ion{Fe}{2}*&    385&  8&  38859&  6\cr
2612.6542&  0.004& \ion{Fe}{2}*&    385&  8&  38660&  8\cr
2626.4511& -0.452& \ion{Fe}{2}*&    385&  8&  38459& 10\cr
2328.1112& -0.684& \ion{Fe}{2}*&    668&  6&  43621&  4\cr
2349.0223& -0.269& \ion{Fe}{2}*&    668&  6&  43239&  6\cr
2381.4887& -0.693& \ion{Fe}{2}*&    668&  6&  42658&  8\cr
2399.9728& -0.148& \ion{Fe}{2}*&    668&  6&  42335&  6\cr
2405.6186&  0.152& \ion{Fe}{2}*&    668&  6&  42237&  8\cr
2607.8664& -0.150& \ion{Fe}{2}*&    668&  6&  39013&  4\cr
2618.3991& -0.519& \ion{Fe}{2}*&    668&  6&  38859&  6\cr
\tablebreak
2632.1081& -0.287& \ion{Fe}{2}*&    668&  6&  38660&  8\cr
2338.7248& -0.445& \ion{Fe}{2}*&    863&  4&  43621&  4\cr
2359.8278& -0.566& \ion{Fe}{2}*&    863&  4&  43239&  6\cr
2405.1638& -0.983& \ion{Fe}{2}*&    863&  4&  42440&  2\cr
2407.3942& -0.228& \ion{Fe}{2}*&    863&  4&  42401&  4\cr
2411.2433& -0.076& \ion{Fe}{2}*&    863&  4&  42335&  6\cr
2614.6051& -0.365& \ion{Fe}{2}*&    863&  4&  39109&  2\cr
2631.8321& -0.281& \ion{Fe}{2}*&    863&  4&  38859&  6\cr
2345.0011& -0.514& \ion{Fe}{2}*&    977&  2&  43621&  4\cr
2411.8023& -0.377& \ion{Fe}{2}*&    977&  2&  42440&  2\cr
2414.0450& -0.455& \ion{Fe}{2}*&    977&  2&  42401&  4\cr
2622.4518& -0.951& \ion{Fe}{2}*&    977&  2&  39109&  2\cr
2629.0777& -0.461& \ion{Fe}{2}*&    977&  2&  39013&  4\cr
2332.00  & -0.720& \ion{Fe}{2}*&   1873& 10&  44754&  8\cr
2348.81  & -0.470& \ion{Fe}{2}*&   1873& 10&  44447&  8\cr
2360.70  & -0.700& \ion{Fe}{2}*&   1873& 10&  44233& 10\cr
2563.30  & -0.050& \ion{Fe}{2}*&   7955&  8&  46967&  6\cr
2715.22  & -0.440& \ion{Fe}{2}*&   7955&  8&  44785&  6\cr
2740.36  &  0.240& \ion{Fe}{2}*&   7955&  8&  44447&  8\cr
2756.56  &  0.380& \ion{Fe}{2}*&   7955&  8&  44233& 10\cr
2166.19  &0.230 &\ion{Ni}{2}*  & 8394 &10  &54557 &10\cr
2217.14  &0.480 &\ion{Ni}{2}*  & 8394 &10  &53496 &12 \cr
2223.61 &-0.140 &\ion{Ni}{2}*  & 8394 &10  &53365 &10\cr
2316.72  &0.268 &\ion{Ni}{2}*  & 8394 &10  &51558 & 8\cr
\enddata   
\tablenotetext{a}{$gf$-values were taken from Kurucz (1995) for
all but the \ion{Fe}{2} lines with wavelengths given more than two
decimal figures. $gf$-values for these come from Morton (2003).} 
\end{deluxetable}
\clearpage

\section{THE MEASURED COLUMN DENSITIES}

In Paper~I, we presented 6.3 hours of VLT/UVES high-resolution
($R \approx 40,000$) spectroscopic observations of QSO~2359--1241 and
identified all the absorption features associated with the outflow
emanating from this object. The unprecedented high signal-to-noise data
from five unblended troughs of \ion{Fe}{2} resonance lines yielded tight
constraints on outflow trough formation models.

As expected we found that the apparent optical depth model
($\tau_{ap}\equiv-\ln(I)$, where $I$ is the residual intensity in the
trough) gives a very poor fit to the data and greatly underestimates the
ionic column density measurements. We found that a power-law distribution
model for absorption material in front of the emission source gives a
better fit to the \ion{Fe}{2} data than does the standard partial covering
factor model (see Sections 1, 3, 4, and Fig.\ 4 of Paper~I). The power-law
distribution model is one in which the outflow fully covers the source,
but does so inhomogeneously. This has the characteristic of allowing for
non-black saturation of absorption troughs at a large source distance
($\sim1$~kpc; see de~Kool et~al.\ 2002c and Arav et~al.\ 2005 for more
detailed investigations of inhomogeneous source coverage). Physically,
this requires the outflow to contain many ``cloudlets'' with dimensions
smaller than the physical span of the source. The inhomogeneous nature
of the absorption also allows the possibility of structure at the bottoms
of otherwise ``saturated'' absorption troughs. Important for the physical
conditions analysis of the present paper was the finding that the partial
coverage and inhomogeneous absorption models yield similar column density
estimates (see Fig.~6 of Paper~I). This gives us greater confidence in the
derived column density values, as they are somewhat model-independent (the
reasons both methods yield similar estimates are described in Paper~I).

Finally, in Paper~I we concluded that the power-law distribution model
is more physically plausible than the partial covering model for outflow
such as this one. We thus used the power-law model as presented there to
extract the column densities of all ions and ionic energy levels present
in the data. For consistency we used the power-law exponent as a function
of velocity of the $E=0$ \ion{Fe}{2} lines for all other \ion{Fe}{2}
energy level and for troughs from all other ions. This assumption should
be robust for the \ion{Fe}{2} troughs, but perhaps less so for the troughs
of other ions. We direct the reader to Paper~I for details concerning this
column density extraction model. In Table~1 we provide the identifications
for all transitions observed in outflow troughs. In Table~2 we provide
the measured column densities of all species identified for the strongest
absorption component, {\bf e} (see Paper~I), many of which will be used to
constrain the photoionization models, as we discuss in the next section.

%%%
\clearpage
\begin{deluxetable}{lrcc}
\tabletypesize{\scriptsize}
\tablecaption{Measured and model predicted column densities for the
major outflow trough in QSO~2359--1412.}
\tablewidth{0pt}
\tablehead{
\colhead{Species} & \colhead{E(cm$^{-1}$)}& 
\multicolumn{2}{c}{{$\log_{10}$ N (cm$^{-2}$)}} 
 \\
 & & \colhead{Observed} & \colhead{Model\tablenotemark{a}} 
}
\startdata 
\ion{H}{1}  &&  &18.59\\
\ion{He}{1}*&& 14.14$\pm$0.03 &14.11\\ 
\ion{He}{1} &Total  & &17.14 \\
\ion{He}{2} &Total  & &19.46 \\
\ion{Fe}{2} &     0& 13.86$\pm$0.02 &13.84 \\ 
\ion{Fe}{2} & 385  & 13.51$\pm$0.04 &13.48 \\
\ion{Fe}{2} & 668  & 13.26$\pm$0.03 &13.28 \\
\ion{Fe}{2} & 863  & 13.06$\pm$0.03 &13.08 \\
\ion{Fe}{2} & 977  & 12.85$\pm$0.03 &12.79 \\
\ion{Fe}{2} &1873   & 13.89$\pm$0.05\tablenotemark{b} &13.83 \\
\ion{Fe}{2} &7955   & 12.70$\pm$0.04 &12.67 \\
\ion{Fe}{2} &Total  & &14.41 \\
\ion{Mg}{1} &Total  & 11.92$\pm$0.03 &13.01 \\
\ion{Mg}{2} &Total  & $>$13.81 &15.22 \\  
\ion{Si}{2} &  287  & 14.9$\pm$0.1 &15.19 \\
\ion{Si}{2} &Total  & 14.9$\pm$0.1 &15.38 \\ 
\ion{Al}{3} &Total  & $>$13.9 &14.35 \\
\ion{Ca}{2} &Total  & 12.52$\pm$0.03 &12.90 \\
\ion{Mn}{2} &Total  & 12.71$\pm$0.03 &12.26 \\
\ion{Ni}{2} &   0 & &13.50\tablenotemark{c} \\
\ion{Ni}{2} &8394   & 12.79$\pm$0.04 &12.98\tablenotemark{c} \\ 
\ion{Ni}{2} &Total  & &13.70 \\
\enddata
\tablenotetext{a}{Predicted column densities are from an optimal
photoionization model with: MF87 SED, solar abundances, constant gas
density $\log n_H(\rm{cm}^{-3}) = 4.4$, $\log N_H(\rm{cm}^{-2}) = 20.556$,
$\log U_H = -2.418$.}
\tablenotetext{b}{Trough partially contaminated with those of other
\ion{Fe}{2} transitions.}
\tablenotetext{c}{Calculated from the \ion{Ni}{2} model of Bautista
(2004) assuming pure collisional excitation.}
\end{deluxetable} 
\clearpage

\section{PHOTOIONIZATION MODELS}

\subsection{General Methodology}

The observed spectrum of QSO~2359--1241 is rich in absorption lines from
singly ionized species, such as \ion{Mg}{2}, \ion{Si}{2} and \ion{Ca}{2},
and notably from iron-peak species. We determine the physical conditions
within the main absorbing component centered on --1376 km~s$^{-1}$
(component {\bf e}; see Paper~I) and integrated over the range in radial
velocity from --1320 km~s$^{-1}$ to --1453 km~s$^{-1}$. Our analysis of
the spectrum is mostly based on the column density in the metastable
2~$^3S$ excited state of \ion{He}{1} (hereafter, \ion{He}{1}$^*$)
and the column densities of \ion{Fe}{2}. For \ion{Fe}{2} we have the
level-specific column densities for the ground level ($a~^6$D$_{9/2}$)
as well as for the excited levels $a~^6$D$_{7/2}$ at 385 cm$^{-1}$,
$a~^6$D$_{5/2}$ at 668 cm$^{-1}$, $a~^6$D$_{3/2}$ at 863 cm$^{-1}$,
and $a~^6$D$_{1/2}$ at 977 cm$^{-1}$ within the ground term, and the
$a~^4$D$_{7/2}$ level at 7955 cm$^{-1}$ within the second excited
term. The total column density in \ion{Fe}{2} is estimated by scaling
from the sum of the above levels (see below). These column densities
as measured from our VLT spectra are presented in the upper half of
Table~2. Together they constrain the physical conditions of the outflow
near its hydrogen ionization front (where the bulk of \ion{He}{1}$^*$
and \ion{Fe}{2} form), as well as its total column density, as we show
below. Moreover, modeling the ionization structure of these two species
combined is secure because the photoionization and recombination cross
sections for hydrogen and helium are well known and the ionization
fraction of \ion{Fe}{2} near the ionization front of the cloud closely
tracks that of H$^+$ due to charge exchange reactions. Table~2 also
lists the measured column densities of several other ions identified in
the spectra.

The diagnostic power of the combined \ion{He}{1}$^*$ and \ion{Fe}{2} lines
results from their different responses to temperature and density. The
\ion{He}{1}$^*$ column density depends on the population of the 2~$^3S$
level of \ion{He}{1}, which is populated by recombination from \ion{He}{2}
and varies with temperature, in the sense that the lower the temperature
the higher the recombination rate. An additional mechanism to populate
the 2~$^3S$ occurs through thermalization by electronic collisions with
the very long-lived 2~$^1S$ level. The main de-population mechanisms of
the 2~$^3S$ level are magnetic dipole radiative decay to the ground level
and electron impact excitation to the neighboring 2~$^1S$ (the dominant
channel) and $2~^3P$ levels. For $T < 15,000$~K collisional ionization can
be neglected as a depopulation mechanism of this level (see Clegg 1987).
The net result is that the population of the 2~$^3S$ level becomes
nearly independent of density for electron densities substantially above
the critical density $\approx 3-4 \times 10^3$~cm$^{-3}$ (Osterbrock
\& Ferland 2006; see also Arav et~al.\ 2001). The column density in
\ion{He}{1}$^*$ is then set by that in \ion{He}{2}, which in turn is set
by the ionization parameter for a fixed spectral energy distribution of
the incident continuum.

The \ion{Fe}{2} lines in our spectrum arise from the ground and first
excited multiplets of the ion. The populations of these levels are
dominated by electron impact excitation, thus the populations depend
approximately linearly on electron density for densities up to $\sim
10^5$~cm$^{-3}$. The populations, particularly that of the $a~^4D_{7/2}$
at 7955 cm$^{-1}$ level, also increase monotonically with temperature
under collisional excitation conditions.
 
For the present work we use the photoionization modeling code Cloudy
(v06.02; Ferland et~al.\ 1998) to compute spectral models of the
outflowing absorbing gas in QSO~2359--1241. This version of {\sc
cloudy} includes the 371-level Fe$^+$ model atom of Verner et al.\
(1999), as well as the detailed model \ion{He}{1} atom of Porter et~al.\
(2005). The default version of the model \ion{He}{1} atom (nLS-resolved
up through principle quantum number n=6, plus 20 additional LS-collapsed
levels lying above) was determined to be sufficient for our purposes. We
assume constant total hydrogen density ``clouds'' of solar abundances
(in particular $\log \rm{(He/H)} = -1.0$ and $\log \rm{(Fe/H)}
= -4.5498$ by number; \citep{holweger}, and adopted the Mathews \&
Ferland (1987) quasar spectral energy distribution (hereafter, the MF87
SED) as the incident continuum spectrum. The solar abundances are from
\cite{allende02,allende01} for C and O, \cite{holweger} for N, Ne, Mg,
Si, and Fe, and \cite{grevesse} for the rest.)

\subsection{Some General Physical Considerations}

Before we proceed to determine the physical conditions within the outflow
of QSO~2359--1241, let us first examine the formation of \ion{Fe}{2}
and \ion{He}{1}$^*$ within a photoionized cloud. Figure~1a shows the
ionic fractions of \ion{Fe}{2} and of helium in the \ion{He}{1}* state
within a representative photoionized gas cloud. Note that we plot the
\ion{He}{1}* fraction relative to that of the total iron abundance
so that the comparison of the curves for \ion{He}{1}($2~^3$S) and
\ion{Fe}{2} is then independent of the iron abundance. The conditions of
this model cloud are: hydrogen number density, $\vy{n}{H} = 10^{4.4}$
cm$^{-3}$, ionization parameter, given by $\log \vy{U}{H} = -2.418$,
solar abundances, and the MF87 SED. The particular choice of $\vy{n}{H}$
and $\vy{U}{H}$ will become apparent below. Here, the ionization parameter
$\vy{U}{H} \equiv \Phi_H/c\vy{n}{H}$, where $\Phi_H$ (cm$^{-2}$~s$^{-1}$)
is the hydrogen ionizing photon flux and $\vy{n}{H}$ (cm$^{-3}$) is the
total hydrogen number density. The cloud is bounded by a total hydrogen
column density $\log N_H$(\rm{cm}$^{-2}) = 21$. Note that the column
densities of \ion{Fe}{2} and \ion{He}{1}$^*$ vary rapidly approaching
the hydrogen ionization front (at $\log N_H(\rm{cm}^{-2}) \approx 20.6$
in this model). In particular, the \ion{Fe}{2}/Fe ratio rises $\sim$~4
orders of magnitude to values approaching unity within \ion{He}{2} zone
(as indicated by the bump in \ion{He}{1}($2~^3$S)), with most of this
change occurring just inside the hydrogen ionization front. For the
cloud shown, 90\% of the \ion{Fe}{2} column density forms within 20\%
of the cloud volume lying to the left of the vertical dashed line. By
comparison the \ion{He}{1}$^*$ column density increases by just 25\%
over the same volume of the cloud. Since nearly all of the observable
\ion{Fe}{2} forms right along the hydrogen ionization front, the model
results will depend only weakly on the iron abundance.

\clearpage
\begin{figure}
\rotatebox{0}{\resizebox{\hsize}{\hsize}
{\plotone{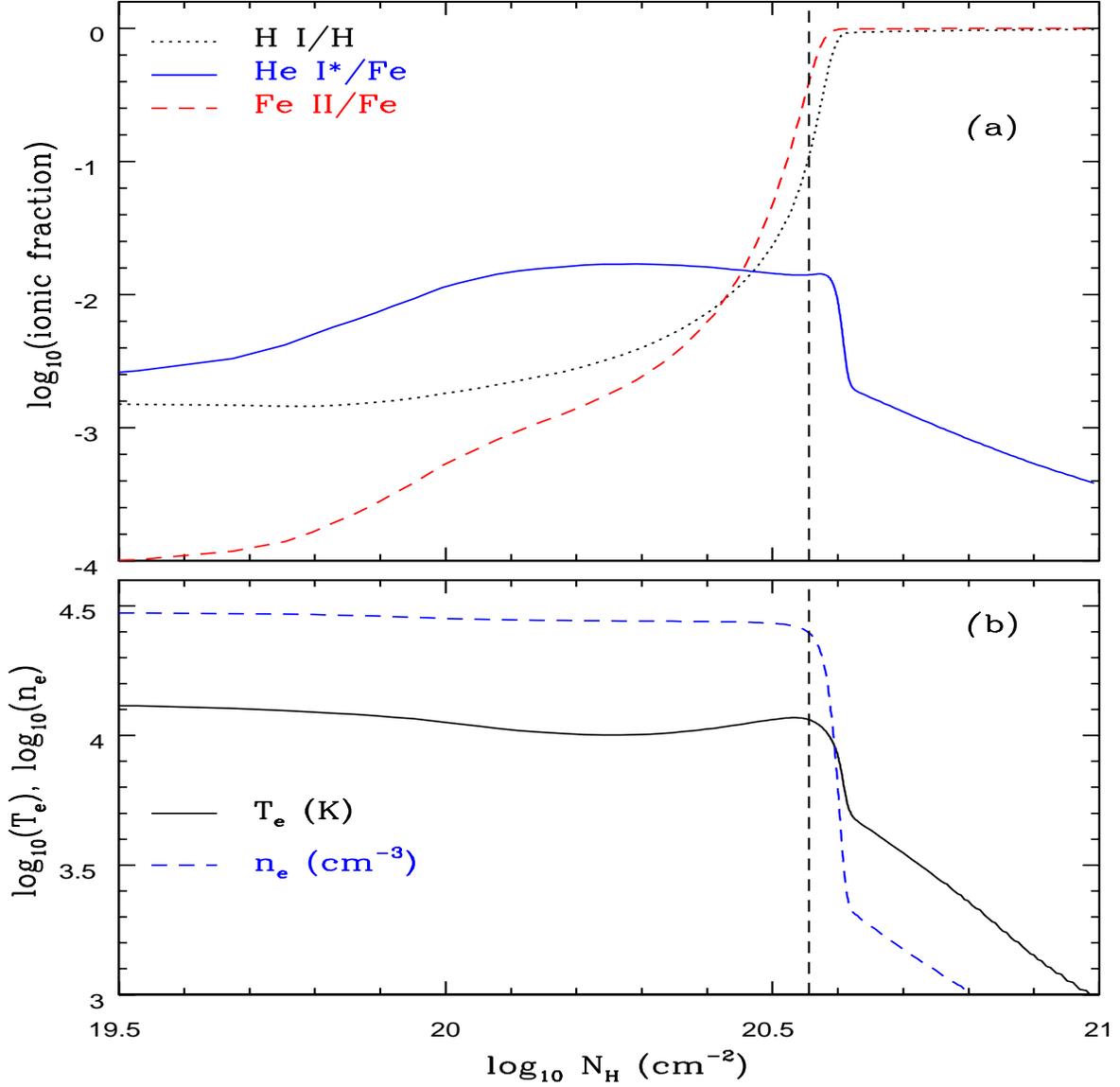}}}
\caption{Physical structure of a photoionized cloud vs.\ total hydrogen
column density. The parameters of the model are given in the text. The
upper panel shows the fraction of neutral helium in the $2~^3S$ level
and the ionic fractions of \ion{H}{1}, \ion{Fe}{2}. The \ion{He}{1}*
fraction has been normalized to the total iron abundance. The lower panel
depicts the behavior of the electron density and electron temperature.
The vertical dashed line indicates the total hydrogen column density
of the model to match the observed column densities of \ion{Fe}{2}
and \ion{He}{1}* in QSO~2359--1241. See text for details.}
\label{modelcond} 
\end{figure}
\clearpage

Figure~1a also shows that the \ion{Fe}{2}/Fe ratio remains near
unity behind (in the more neutral side of) the ionization front. As
a consequence, the integrated \ion{Fe}{2}/\ion{He}{1}$^*$ ratio
continues to grow in clouds with column densities that extend beyond
the hydrogen ionization front, whose column density scales as $N_{ion}
\approx 10^{23}U_H$ cm$^{-2}$. This is similarly the case for other
singly-ionized species of atoms with low first ionization potential ($<
13.6$~eV) that we measure, such as \ion{Mg}{2}, \ion{Si}{2}, \ion{Ca}{2},
\ion{Mn}{2}, and \ion{Ni}{2}. However, as we show in the next section, the
observed \ion{Fe}{2} column density in QSO~2359--1241 is largely formed
in the warm region on the ionized side of the hydrogen ionization front.

In Figure~1b we show the run of electron temperature and electron density
through the same representative cloud. Note the sharp drop in both
quantities behind the hydrogen ionization front. These are accompanied by
a corresponding rapid depopulating of excited states of the \ion{Fe}{2}
ion, even within the ground a~$^6$D term, but especially those in the
a~$^4$D (e.g., the 7955 cm$^{-1}$ level) which lie nearly 1~eV above the
ground. Thus, if absorption occurred within the ``cold'' neutral zone,
the \ion{Fe}{2} column density would be much greater and dominated by the
resonance lines out of the ground term (especially those those arising
from the 0.00 cm$^{-1}$ level), unlike that which is observed in the
spectrum of QSO~J2359--1241. We discuss this further in the next section.

Figures 1a,b depict general qualitative physical properties of these
types of absorption line systems, while also providing important physical
constraints on our photoionization models. Furthermore, the particular
model shown was found to best reproduce the observed \ion{Fe}{2} and
\ion{He}{1}* column densities. The model and the optimization technique
employed to match the measured column densities are described in detail
in the following section.

%%%%%%%%%%%%%%%%%%%%%%

%%%%%%%%%%%%%%%%%

\subsection{Photoionization Models of the Outflow in QSO~2359--1241}

Before attempting self-consistent photoionization modeling of the
absorbing gas it is useful to constrain the physical conditions of the
ionized region based on the observed level specific column densities of
the \ion{Fe}{2} ion as well as that of \ion{He}{1}$^*$ (see Table~2). We
use a stand-alone spectral model of \ion{Fe}{2} \citep{baupra98} to
investigate the populations of the a~$^6D_j$ and a~$^4D_j$ levels as
functions of {\em fixed values in electron density and temperature}. We
find that the observed relative column densities are consistent with
$\log(n_e)$ = $4.4\pm0.1$ and $T_e> 9000$~K. The former is constrained
by the relative populations within the a~$^6D_j$ ground term, while the
latter (at this density) is constrained by the a~$^4D$ 7955 cm$^{-1}$
population. Further, we find no evidence for significant variations
of physical conditions along the radial velocity space covered by the
\ion{Fe}{2} absorption troughs. In Figure~2 one sees that across a span
of $\sim100$~km~s$^{-1}$ the the column densities of the levels relative
to the 0.00 cm$^{-1}$ ground level are constant within the error bars,
as also demonstrated by linear fits to all three curves. This implies
that $n_e$ is similarly constant across this outflow component. As seen
in Figure~1 the photoionization model solution is reached (vertical
dashed line) before the values of $n_e$ and $T_e$ begin plummeting due
to the \ion{H}{1} ionization front. If the absorption had taken place
deeper within or behind the ionization front, we should have then seen
a rapid decline of the N(\ion{Fe}{2}*)/N(\ion{Fe}{2}) at either high or
low velocity, assuming the velocity of the flow changes monotonically
with radius (see the discussion regarding the physical nature of the
outflow in Section~4 of Paper~I). This is because the relative level
populations within a~$^6D_j$ term are roughly linearly dependent on $n_e$
for $n_e$ well below the critical densities of the excited levels, and
that of the 7955 cm$^{-1}$ level also depends on $T_e$ (primarily via
the Boltzmann factor) in our temperature range. Both effects work in
the same direction: the lowering of $n_e$ and $T_e$, which occurs deep
within and beyond the hydrogen ionization front (see Figure~1), results
in significantly reduced populations of the excited levels, as mentioned
in Section~3.1. For example, in the range of electron density $\log(n_e)
= 3.4 - 4.4$ the excited state level populations within the a~$^6D_j$
term fall by about a factor of 5, while that of the 7955 cm$^{-1}$
level falls by a decade due to electron density alone and a factor of
$\sim$2 as the temperature falls from 11,500 K to half that.

>From these preliminary investigations we conclude that the observed
\ion{Fe}{2} troughs form mostly within a narrow region of the outflow
near the hydrogen ionization front for which the electron density and
temperature do not vary significantly, and that a hydrogen ionization
front does not fully form within the flow (see Figure~1). This
then constrains the cloud column density to be $\log N_H < 20.6 +
\log(U_H/10^{-2.4})$). This is in contrast to the findings of de~Kool
et~al.\ (2001; 2002a,b) in their analysis of intrinsic \ion{Fe}{2}
troughs in other quasars. This is an important conclusion, since the
rapidly plunging values of electron density and temperature behind the
hydrogen ionization front would make for more difficult modeling.

%%%
\clearpage
\begin{figure}
\rotatebox{90}{\resizebox{\hsize}{\hsize}
{\plotone{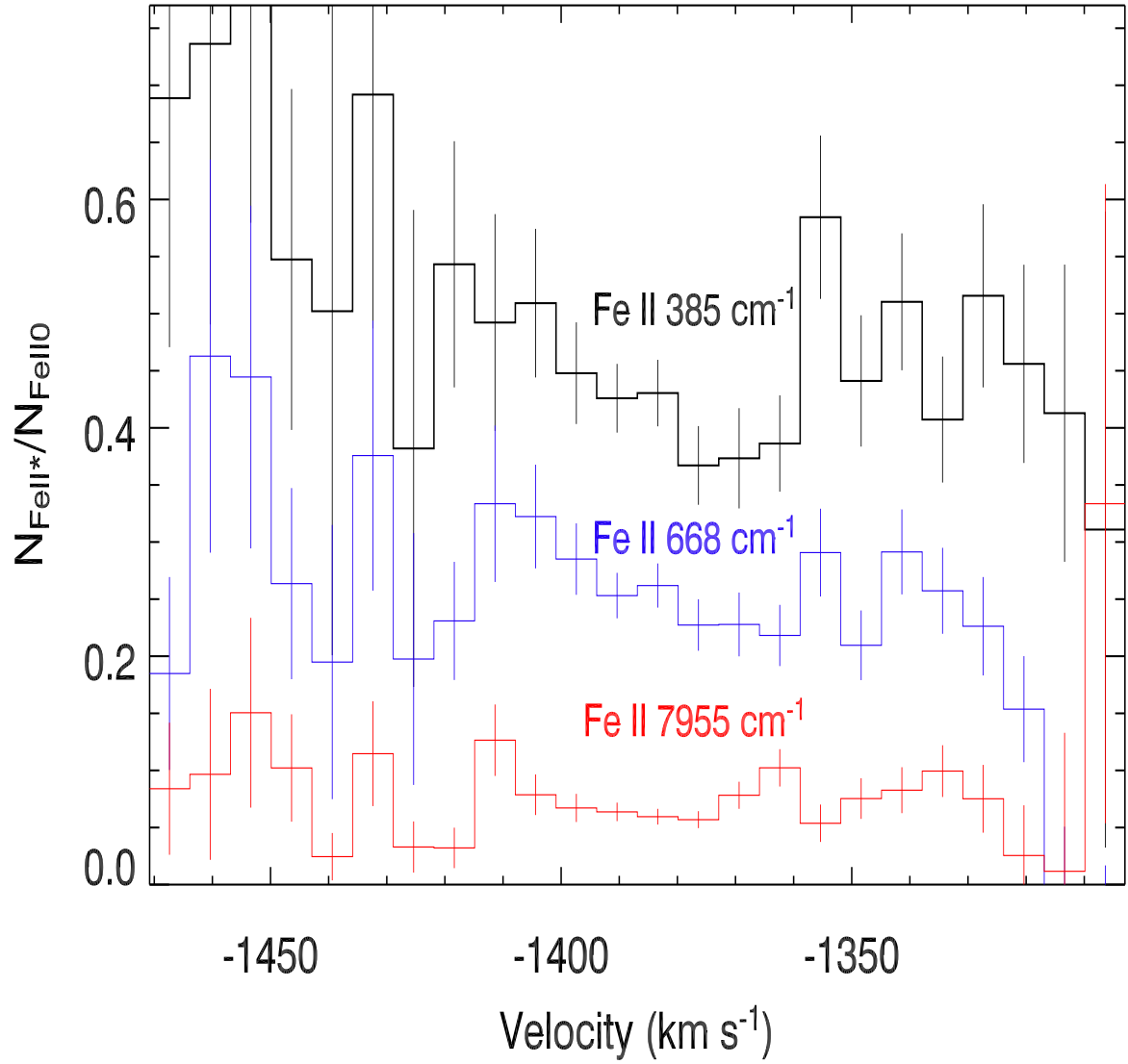}}}
\caption{Ratios of observed column densities of excited \ion{Fe}{2}
levels to the ground level vs.\ velocity shift along the trough.}
\label{velratios}
\end{figure}
\clearpage

We also use the analytic formula of Clegg (1987; see also Equation~3
of Arav et~al.\ 2001) to estimate a ratio $n(2^3S)/n($He$^+$) $\approx
5 \times 10^{-6}$ for the above electron density and a temperature
of $10^4$~K. Further, if we assume a solar (He/H) abundance and that
(He$^+$/He) $\sim$ 0.8 for the ionized portion of the cloud (as is
typical for standard AGN SEDs), we estimate a total hydrogen column
density of $\vy{N}{H} \sim 2.5 \times 10^6 N(2^3S) \sim 10^{20.5}$
cm$^{-2}$, given our measurement of $N(2^3S)$ quoted in Table~2. The
substantial improvement in quality of the spectral data of QSO~2359--1241
(see also Paper~I) over that presented in Arav et~al.\ (2001) has thus
already yielded much stronger constraints in the physical conditions
within the outflow. We next proceed with the more detailed modeling.

In building a fully self-consistent photoionization model of the outflow
of QSO~2359--1241 based on the intrinsic absorption lines, we adopt a
two-step iterative photoionization modeling procedure. Using the default
parameter optimization scheme built into Cloudy, the ionization
parameter $\vy{U}{H}$ and total hydrogen column density $\vy{N}{H}$ are
optimized to simultaneously match the total column density in \ion{Fe}{2}
and the column density in the excited state \ion{He}{1}$^*$.

In the first pass we adopt a hydrogen number density, $\vy{n}{H}$,
equal to the electron density derived from the \ion{Fe}{2} level specific
column densities ($10^{4.4}$~cm$^{-3}$), and the total \ion{Fe}{2} column
density is estimated from the sum of the level specific column densities
determined from the observations and the results of our stand-alone
\ion{Fe}{2} model atom for the above density and a temperature of
10,000~K. The Cloudy optimizer minimizes $\chi^2$ between the
computed and measured target column densities (with assumed equal
weighting) for various values of $\vy{U}{H}$ and $\vy{N}{H}$.  To speed up
computations and since we are not concerned with the detailed \ion{Fe}{2}
level populations during this step, the default 16-level model atom of
Fe$^+$ within Cloudy is utilized during the optimization. This
subset of the full 371-level model atom includes all levels up through
a~$^4$P near 13,500 cm$^{-1}$, i.e., the four lowest terms of the
ion. For a model of this size only electron impact excitation followed
by radiative decay through forbidden transitions is considered in the
excitation. However, as verified later, radiative processes contribute
little to the excitation of the \ion{Fe}{2} levels observed.

Once a solution is found, we update the value of the total \ion{Fe}{2}
column density (this converges to be 1.80 times the sum over the six
most reliably determined level-specific column densities), and then
fine-tune the value of $\vy{n}{H}$ by computing a grid in total hydrogen
number density, spanning a decade to either side of the starting density
in 0.05 dex steps, for fixed values in $U_H$ and $N_H$ as determined
during the previous optimization step. To provide sufficient accuracy
in the predictions of the level populations under study, the Cloudy
models computed in the gas density grid use a 99-level model atom of
Fe$^+$ (a subset of the full 371-level model atom), which includes all
levels up through 50,212.8 cm$^{-1}$. This model atom accounts for the
dominant channels for photoexcitation by continuum radiation and other
processes. Turning on the full 371-level model atom, that accounts for
\ion{H}{1} Lyman $\alpha$ fluorescence and other processes, has virtually
no impact on the level populations relevant to our study. By comparing the
predicted \ion{Fe}{2} column densities for all levels observed with the
measured values in Table~2, we are able to choose the most appropriate
value in total hydrogen gas number density $\vy{n}{H}$. This procedure
converges rapidly to a final solution. The final fit parameters are $\log
\vy{n}{H} = 4.4$ and $\log \vy{U}{H}= -2.418$, which were used for the
modeled cloud shown in Figure~1, and $\log \vy{N}{H}= 20.556$, which
is indicated by the vertical dashed line in that figure. As a further
point of interest, this model's electron density-weighted temperature
in the Fe$^+$ zone is $11,500$~K.

%%%
\clearpage
\begin{figure}
\rotatebox{-90}{\resizebox{\hsize}{\hsize}
{\plotone{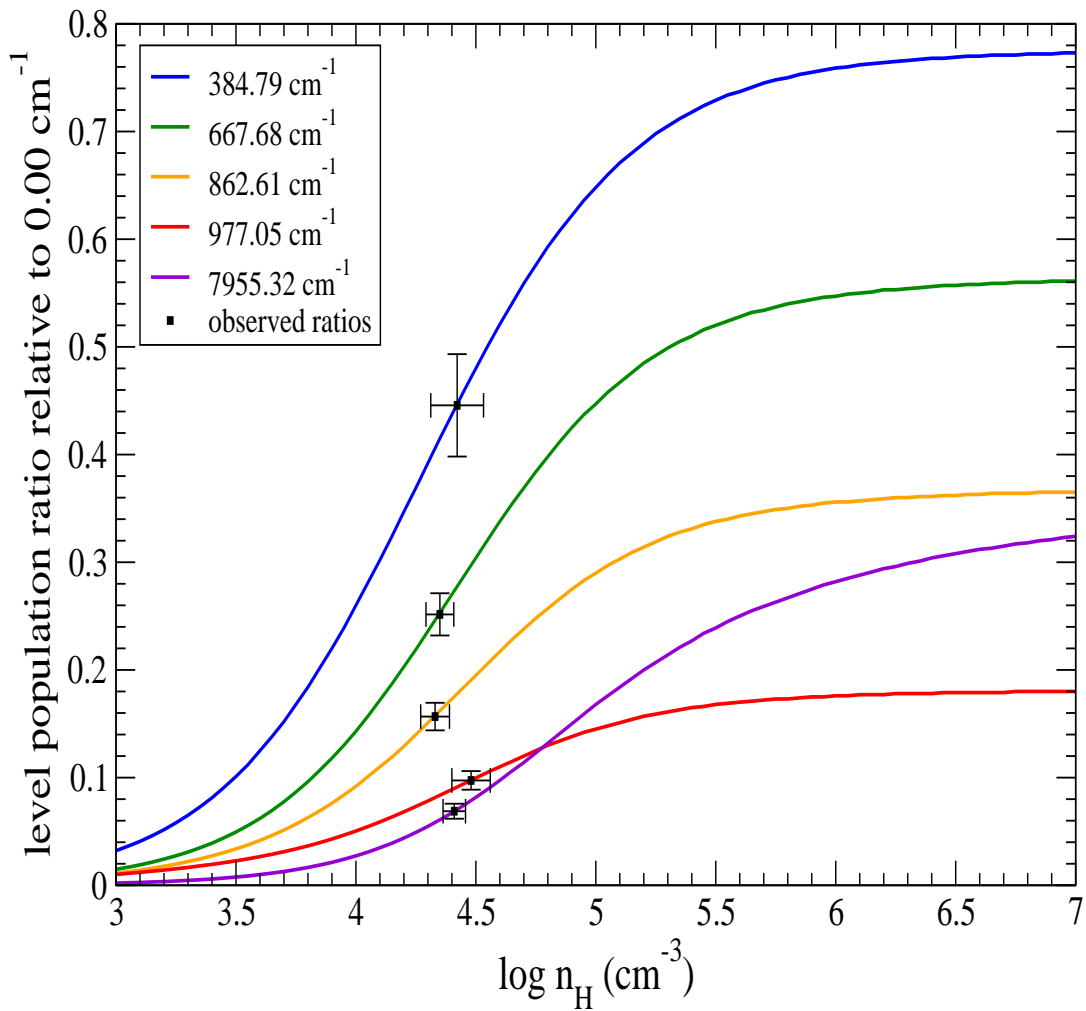}}}
\caption{Calculated level populations of excited levels of \ion{Fe}{2}
relative to the ground level vs.\ hydrogen number density in the
photoionization model. The points indicate the observed column densities
of each level relative to the ground level. See text for details.
All five independent measurements are consistent with $\log n_H = 4.4$
with less than 20\% dispersion.}
\label{levelpop}   
\end{figure}
\clearpage

In Figure~3 we illustrate the populations of various \ion{Fe}{2}
levels relative to the ground level as a function of $\vy{n}{H}$,
assuming the final optimized values of $U_H$ and $N_H$ provided in
the previous paragraph. We have ascertained that none of the models
presented in Figure~3 have fully-formed hydrogen ionization fronts
within them, and so the relative level populations are indicative of
the more nearly constant conditions in $n_e$ and $T_e$ self-consistently
computed within the ionized portion of the cloud \footnote{The lower gas
density clouds will be slightly cooler, all else being equal, due to the
increased efficiency in the forbidden emission line cooling. The hydrogen
ionization front will then occur at slightly lower column densities
due to the increased recombination rate at lower temperature. Thus a
larger {\em total} \ion{Fe}{2} column density would be predicted for a
fixed value in the cloud's total hydrogen column density. These small
changes in temperature with gas density affect mainly the column density
in the 0.00~cm$^{-1}$ level as well as the relative population in the
7955~cm$^{-1}$ level in Figure~3.}.

There is very good agreement between individual determinations of the
gas number density ($\log \vy{n}{H} \approx 4.4$), based on the ratios
of the measured \ion{Fe}{2} level specific column densities (points with
error bars). The 1-sigma uncertainties in the ratios are statistical
only, while those in the number density are estimated by considering the
range in density that results from the spread in the level ratio error
bar in each direction. The robustness in the determination of the gas
density is also indicated by the excellent matches between the observed
and predicted level specific \ion{Fe}{2} column densities presented in
Table~2. Altogether, these results indicate that $n_H$ remains relatively
stable through the outflow, at least within the \ion{Fe}{2} zone.

For completeness we also list in Table~2 the model predictions for
all species measured in the spectra. We made no attempt to constrain
the models based on these ions' measurements. With the exception of
\ion{Si}{2}, there are significant differences between predictions and
observations for these species, e.g., \ion{Mg}{1} and \ion{Ca}{2}. In
part these could be due to current uncertainties in their ionization
balance, largely owing to errors in the dielectronic recombination rate
coefficients relevant to these heavy, low ionization species (see for
example, Section~4.3 of Ferland et~al.\ 1998 for a recent review, but see
also Chakravorty et~al.\ 2008 which discusses recent progress in improving
these rate coefficients). \ion{Fe}{2} is largely immune from this problem,
as mentioned in Section~3.1, because of its strong coupling to the
hydrogen ionization balance via charge transfer. On the other hand, the
differences between theory and observation are also suggestive of chemical
abundances that depart from solar values. In fact super-solar metalicities
are expected for gas whose origin is the center of a quasar host galaxy
(see Hamann et~al.\ 2007 for a recent review). Nearly solar (Si/Fe) and
sub-solar (Mg/Fe) and (Ca/Fe) together are consistent with the presence
of super-solar gas metalicities in galactic chemical evolution models
of massive galaxies (see, for example, Ballero et~al.\ 2008), although
these are not likely to explain the full discrepancies. Unfortunately,
too few observational constraints as well as uncertainties in the atomic
data for these third and, especially, fourth row elements (Ca, Mn)
do not allow us to draw definitive conclusions at the present time.

\subsection{Additional considerations: alternative metalicities and SEDs}

We have found a good model for the outflow in QSO~2359--1241 by
self-consistent optimization of the total hydrogen gas density, ionization
parameter and total hydrogen column density. In particular, for reasons
already provided we believe our derived value in the hydrogen gas number
density is robust. However, the model contains two basic assumptions
concerning the gas elemental abundances (solar) and the adopted SED
(MF87), neither of which are well-constrained by the observations. Each
of these assumptions independently can have important effects on the
physical conditions and so on some of the inferred physical parameters
such as ionization parameter and total hydrogen column density, but
in at least some cases can cancel each other under the constraints of
our observations.

The assumed metalicity of the cloud affects the thermal structure
of the cloud, because metals as a whole provide radiative cooling to
the plasma. Thus, for a fixed SED, super-solar metalicity clouds are
cooler than our optimized model, and sub-solar metalicity clouds would
be warmer. Thus, increasing the gas metalicity (along with Fe/H) also
decreases the column density of the ionized zone, all else being equal, as
the lower electron temperatures result in more rapid recombination rates
and as the heavy elements become increasingly important opacity sources.
However, we find that models with metalicities exceeding 2--3$\times$
solar, for a fixed MF87 SED, are too cool ($T$ falls below 9000~K) to
reproduce the observed column density ratio between the 7955~cm$^{-1}$
and the ground 0.00~cm$^{-1}$ levels. For larger gas metalicities, the
incident SED must be correspondingly harder than the MF87 SED to push
back up the electron temperature within the ionized zone. The observations
of excited \ion{Fe}{2} offer good constraints on the minimum temperature
near the ionization front ($T > 9000$~K), but provide no upper limit to
the temperature below 20,000~K.

The shape of the ionizing SED also affects the temperature structure
of the cloud, as well as the sharpness of onset of the ionization
front. The distribution of hydrogen and helium ionizing photons
determines the temperature of the plasma at the illuminated face, and
then the distribution of the EUV and X-ray photons determines how sharply
temperature and ionization drop at the ionization front and beyond. A
hard SED, i.e., with a large fraction of photons in the EUV and X-rays,
leads to an extended ionization front, while a soft SED, with relatively
more photons concentrated near 1~Rydberg, results in sharper declines in
temperature, ionization, and electron density. We find that gas of solar
abundances illuminated by an SED with a logarithmic X-ray to UV flux ratio
$\alpha_{ox} < -1.7$ has electron temperatures that are too low near the
hydrogen ionization front to explain the observed \ion{Fe}{2} column
densities. Based on the high optical-UV luminosity of QSO~2359--1241
(Brotherton et~al.\ 2001) and the empirical relation between $\alpha_{ox}$
and $L_{uv}$ in quasars (e.g., see Strateva et~al.\ 2005), an $\alpha_{ox}
\approx -1.6 \pm 0.2$ might be expected. What little we do know about the
X-rays in QSO~2359--1241 comes from {\em Chandra} observations of modest
quality (16 photons detected) that find $\alpha_{ox} \approx -1.4 \pm 0.1$
(Brotherton et~al.\ 2005 and Brotherton 2008, private communication;
estimated error bar is statistical only), consistent with the above
empirical relation and more importantly is the same value as that in our
assumed SED (MF87). We refer the reader to Arav et~al.\ (2001) and to
Brotherton et~al.\ (2001) for information pertaining to this object's
rest frame opt-UV spectrum.

In conclusion, the few observational constraints available do support
an incident continuum SED whose average ionizing photon energy is
similar to the MF87 spectrum, although there remains the possibility
(we consider unlikely) that it is substantially harder. Furthermore,
the gas metalicity is unlikely to be severely sub-solar. Therefore,
the total column density of the absorbing gas as derived in the previous
section seems secure to within a factor of 2--3.

\section{Discussion and Conclusions}  

First, we compare the results of this investigation to the analysis of
the Keck/HIRES observations of QSO~2359--1214 by Arav et al.\ (2001;
hereafter HIRES paper). Using the same MF87 SED, the HIRES analysis found
$\log N_H = 20.2$ and $\log U_H = -2.7$, compared to $\log N_H = 20.6$
and $\log U_H = -2.4$ for the VLT analysis. These factors of $\sim2$
differences are mainly attributed to using apparent optical depth
methods to extract the \ion{Fe}{2} and \ion{He}{1}* column densities.
As pointed out in the HIRES paper, the data was not of high enough
signal-to-noise ratio to permit more sophisticated analyses.  Even so,
the HIRES paper already showed that the outflow is not shielded by a
hydrogen ionization front, a result confirmed by the VLT analysis.

The important leap in diagnostic power for the VLT data came from the
ability to accurately measure the population levels of the excited
\ion{Fe}{2} levels, allowing us to pin point the number density
of the outflow to $\log(n_e)=10^{4.4}$ cm$^{-3}$ to better than
20\% accuracy. This is both qualitatively and quantitatively a great
improvement over the lower limit of $\log(n_e)=10^{5}$ cm$^{-3}$ available
from the HIRES data. This result is not accidental. The main reason we
invested 6.5 hours of VLT observation on this outflow was precisely
to yield a data set from which an accurate $n_e$ could be extracted.
This determination of $n_e$ will allow us to determine the distance of
the outflow from the central source and thus measure its mass flux and
kinetic luminosity. This demonstrates the importance of taking high
quality spectra of such outflows.

Other important results arising from the measured populations of
the \ion{Fe}{2} levels are the determination of a lower limit to the
temperature of the \ion{Fe}{2} region and the realization that the
absorption spectrum forms before the hydrogen ionization front, beyond
which the temperature and ionization drop sharply. The temperature
determination was crucial in constraining a whole family of SEDs
and gas metalicities that would yield very different temperatures
in the \ion{Fe}{2} region, and consequently allowing for a more
secure determination of the total gas column density and ionization
parameter. That the \ion{Fe}{2} absorption occurs in a region of nearly
constant conditions before the hydrogen ionization front is a key to
being able to model the absorption spectrum.

Photoionization modeling allowed us to reproduce quite well the observed
\ion{Fe}{2} and \ion{He}{1}* column densities in the main, {\bf e},
component of the quasar outflow absorption spectrum. Reiterating, we
found $\log N_H \approx 20.6$ and $\log U_H \approx -2.4$. The dominant
error bars to these values come from the uncertainties in the assumed
SED and gas metalicities and come to $\sim$~0.3 dex.

Given the above gas density, ionization parameter, an estimate to an
{\em unobscured} incident bolometric luminosity of $\sim4.7 \times
10^{47}$ ergs~s$^{-1}$ based on the intrinsic reddening correction in
Brotherton et~al.\ (2001, 2005), and a standard cosmology ($H_o = 70$
km~s$^{-1}$~Mpc$^{-1}$, $\Omega_\Lambda = 0.70$, $\Omega_m = 0.30$), we
estimate a distance of component {\bf e} of the outflow from the central
continuum source of $\sim3$ kpc. In a future paper we will similarly
determine the physical conditions in the weaker, lower velocity outflow
components {\bf a--d}, as well as the distances and estimates of the
kinetic luminosities for all components in the outflow, important to
AGN feedback scenarios of galaxy evolution.

\acknowledgements
We acknowledge support from NSF grant number AST~0507772 and from NASA
LTSA grant NAG5-12867. We also would like to thank the anonymous referee
for his or her helpful comments and suggestions.

\clearpage

%%%% References

%%%%%%%%%%%%%%%%%%%%%%

%%%% 

\begin{thebibliography}{20}
\expandafter\ifx\csname natexlab\endcsname\relax\def\natexlab#1{#1}\fi


\bibitem[{{Allende~Prieto} {et al.}(2001)}]{allende01}
Allende Prieto, C., Lambert, D.L., \& Asplund, M.\ 2001, ApJ, 556, L63

\bibitem[{{Allende~Prieto} {et al.}(2002)}]{allende02}
Allende Prieto, C., Lambert, D.L., \& Asplund, M.\ 2002, ApJ, 573, L137

\bibitem[{{Ander} \& {Grevesse}(1989)}]{grevesse}
Anders, E., \& Grevesse, N.\ 1989, Geochim.\ Cosmochim.\ Acta, 53, 210

\bibitem[{{Arav} {et al.}(1997)}]{arav97}
Arav, N.; Barlow, T.A.; Laor, A.; Blandford, R.D.\ 1997, MNRAS 288, 1015

\bibitem[{{Arav} {et al.}(1999a)}]{arab99a}
Arav, N., Becker, R.H., Laurent-Muehleisen, S.A., Gregg, M.D., White,
R.L.; Brotherton, M.S.; de Kool, M.\ 1999a, ApJ 524, 566

\bibitem[{{Arav} {et al.}(2001)}]{arav01a}
Arav, N., Brotherton, M.S., Becker, R.H., Gregg, M.D., White, R.L.,
Price, T., Hack, W.\ 2001a, ApJ 546, 140

\bibitem[{{Arav} {et al.}(2007)}]{arav07}
Arav, N., Gabel, J.R., Korista, K.T., Kaastra, J.S., Kriss, G.A., et~al.\
2007, ApJ 658, 829

\bibitem[{{Arav} {et al.}(2005)}]{arav05}
Arav, N., Kaastra, J., Kriss, G.A., Korista, K.T., Gabel, J., Proga,
D.\ 2005, ApJ 620, 665

\bibitem[{{Arav} {et al.}(2003)}]{arav03}
Arav, N., Kaastra, J.; Steenbrugge, K., Brinkman, B., Edelson, R., Korista, K.T., de Kool, M.\ 2003, ApJ 560, 174

\bibitem[{{Arav} {et al.}(1999b)}]{arav99b}
Arav, N.; Korista, K.T.; de Kool, M.; Junkkarinen, V.T.; Begelman, M.C.\
1999b, ApJ 516, 27

\bibitem[{{Arav} {et al.}(2002)}]{arav02}
Arav, N.; Korista, K.T.; de Kool, M.\ 2002, ApJ 566, 699

\bibitem[{{Arav} {et al.}(2008)}]{arav08} 
Arav, N., Moe, M., Costantini, E., Korista, K.T., Benn, C., Ellison,
S.\ 2008, ApJ (in press)

\bibitem[{{Ballero et al}(2008)}]{ballero} 
Ballero, S.K., Matteucci, F., Ciotti, L., Calura, F., \& Padovani, P.\
2008, \aap, 478, 335

\bibitem[{{Barlow}{et al.}(1997)}]{barlow} 
Barlow, T.A., Hamann, F., \& Sargent, W.L.W. 1997, in ASP Conf. Ser. 128,
Mass Ejection from AGN, ed.\ R.\ Weymann, I.\ Shlosman, \& N.\ Arav
(San Francisco: ASP), 13

\bibitem[{{Bautista}(2004)}]{bautist04}
Bautista, M.A.\ 2004, A\&A 420, 763

\bibitem[{{Bautista} \& {Pradhan}(1998)}]{baupra98}
Bautista, M.A.\ \& Pradhan, A.K.\ 1998, ApJ 492, 650

\bibitem[{{Blandford} \& {Begelmen}(2004)}]{blanford}
Blandford, R, D., Begelman, M. C.\ 2004, MNRAS 349, 68

\bibitem[{{Brotherton et al.}(2001)}]{brother01}
Brotherton, M.S., Arav, N., Becker, R.H., Tran, H.D., Gregg, M.D., White,
R.L., Laurent-Muehleisen, S.A.\ \& Hack, W.\ 2001, ApJ, 546, 134

\bibitem[{{Brotherton et al.}(2006)}]{brother06}
Brotherton, M.S., Laurent-Muehleisen, S.A., Becker, R.H., Gregg, M.D.,
Telis, G., White, R.L., \& Shang, Z.\ 2005, AJ, 130, 2006

\bibitem[{{Cattaneo} {et al.}(2006)}]{cattaneo}
Cattaneo et~al.\ 2005 MNRAS, 364, 407

\reference{} Chakravorty, S., Kembhavi, A.K., Elvis, M., Ferland, G.,
\& Badnell, N.R.\ 2008, \mnras, 384, L24

\bibitem[{{Clegg}(1987)}]{clegg}
Clegg, R.E.S.\ 1987, MNRAS, 229, 31

\bibitem[{{de~Kool} {et al.}(2001)}]{dekool01}
de Kool, M., Arav, N., Becker, R.H., Gregg, M.D., White, R.L., 
Laurent-Muehleisen, S.A., Price, T., \& Korista, K.T.\ 2001, ApJ, 548, 609

\bibitem[{{de~Kool} {et al.}(2002)}]{dekool02a} 
de~Kool, M., Becker, R.H., Gregg, M.D., White, R.L., \&  Arav, N.\ 2002a,
ApJ, 567, 58

\bibitem[{{de~Kool} {et al.}(2002)}]{dekool02b}
de~Kool, M., Becker, R.H., Arav, N., Gregg, M.D., \& White, R.L.\ 2002b,
ApJ, 570, 514

\bibitem[{{de~Kool} {et al.}(2002)}]{dekool02c}
de~Kool, M., Korista, K.T., \& Arav, N.\ 2002c, ApJ, 580, 54

\bibitem[{{Ferland} {et al.}(1998)}]{cloudy}
Ferland, G.J., Korista, K.T., Verner, D.A., Ferguson, J.W., Kingdon, J.B., Verner, E.M.\ 1998, PASP 110, 761 

\bibitem[{{Haiman}(2006)}]{haiman}
Haiman et~al.\ 2006, ApJ, 650, 7

\bibitem[{{Hamann}(2007)}]{hamann}
Hamann, F., Warner, C., Dietrich, M., \& Ferland, G.\ 2007, ASP Conference
Series, volume 373, 653 (astro-ph/0701503)

\bibitem[{{Holweger}(2001)}]{holweger} 
Holweger, H.\ 2001, Joint SOHO/ACE workshop Solar and Galactic
Composition. Ed.\ by Robert, F.  Wimmer-Schweingruber. Publisher: American
Institute of Physics Conference proceedings vol.\ 598 location: Bern,
Switzerland, March 6 - 9, 2001, p.23

\bibitem[{{Hopkins} {et al.}(2006)}]{hopkins}
Hopkins, P.F., Hernquist, L., Cox, T.J., Di Matteo, T., Robertson, B., \& Springel, V.\ 2006, ApJS 163, 1

\bibitem[{{King}(2003)}]{king}
King, A., 2003, ApJ 596, L27

\bibitem[{{Kurucz \& Bell}(1995)}]{kurucz}
Kurucz, K.T. \& Bell, B., Atomic Line Data, Kurucz CD-ROM No. 23. Cambridge,
Mass.: Smithsonian Astrophysical Observatory, 1995

\bibitem[{{Mathews} \& {Ferland}(1987)}]{mf87}
Mathews, W.G., \& Ferland, G.J.\ 1987, ApJ, 323, 456

\bibitem[{{Menci} {et al.}(2006)}]{menci}
Menci, N., et al., 2006, ApJ, 647, 753

\bibitem[{{Morton}(2003)}]{morton}
Morton, D.C.\ 2003, ApJS 149, 205

\bibitem[{{Scannapieco} \& {Oh}(2004)}]{scanna}
Scannapieco, E., Oh, S.P.\ 2004, ApJ 608, 62

\bibitem[{{Silk} \& {Rees}(1998)}]{silk}
Silk, J., Rees, M.J.\ 1998, A\&A 331, L1S

\bibitem[{{Springel} {et al.}(2005)}]{springel}
Springel, V., et al.\ 2005, MNRAS, 361, 776

\bibitem[{{Osterbrock} \& {Fraland}(2006)}]{oster}
Osterbrock, D.E. \& Ferland, G.J. 2006, Astrophysics of Gaseous Nebulae
and Active Galactic Nuclei, 2nd.\ Ed.\ University Science Books, 2006

\bibitem[{{Strateva et al.}(2005)}]{strateva} 
Strateva, I. V., Brandt, W.N., Schneider, D.P., Vanden Berk, D.G., \&
Vignali, C.\ 2005, AJ, 130, 387

\bibitem[{{Porter} {et al.}(2005)}]{porter}
Porter, R.L.; Bauman, R.P.; Ferland, G.J.; MacAdam, K.B.\ 2005, ApJ,
622, 73

\bibitem[{{Telfer} {et al.}(1998)}]{telfer}
Telfer, R.C., Kriss, G.A., Zheng, W., Davidsen, A.F., Green, R.F.\ 1998,
ApJ 509, 132

\bibitem[{{Vernaleo} \& {Reynolds}(2006)}]{vernaleo}
Vernaleo, J. C., \& Reynolds, C. S., 2006, ApJ, 645, 83

\bibitem[{{Verner} {et al.}(1999)}]{verner}
Verner, E.M., Verner, D.A., Korista, K.T., Ferguson, J.W., Hamann, F.,
Ferland, G.J.\ 1999, ApJS 120, 101

\bibitem[{{Wampler}, {Chugai}, \& {Petitjean}(2001)}]{wampler}
Wampler, E. J., Chugai, N. N., \& Petitjean, P.\ 1995, ApJ, 443, 586

\end{thebibliography}
\end{document}